\documentstyle{article}
\begin{document}
\LARGE
\begin{center}
\bf Dimensionality in the Freund-Rubin Cosmology
\vspace*{0.7in}

\large \rm Zhong Chao Wu

Dept. of Physics

Zhejiang University of Technology

Hangzhou 310032, P.R. China

\vspace*{0.55in}
\large
\bf
Abstract
\end{center}
\vspace*{.1in}
\rm
\normalsize
\vspace*{0.1in}

In the $n-$dimensional Freund-Rubin model with an antisymmetric
tensor field of rank $s-1$, the dimension of the external
spacetime we live in must be $min(s, n-s)$. This result is a
generalization of the previous result in the $d=11$ supergravity
case, where $s = 4$.

\vspace*{0.3in}

PACS number(s): 04.65.+c, 11.30.Pb, 04.60.+n, 04.70.Dy

Key words: quantum cosmology, Kaluza-Klein theory, anthropic
principle, dimensionality

\vspace*{0.5in}
e-mail: zcwu@zjut.edu.cn

\pagebreak

\rm

\normalsize

The history of Kaluza-Klein models is almost as long as that of
General Relativity. The idea of dimensional reduction has been
revived many times, for example, in the context of nonabelian
gauge theory, extended supergravity, and  most recently,
$M-$theory or  brane cosmology.

Traditionally, it is assumed that in the Kaluza-Klein models, the
$n-$dimensional spacetime is a product of a $s-$dimensional
manifold $M^s$ and a $n-s-$dimensional manifold $M^{n-s}$. Many
studies have been done to show how to decompose $M$ into the
product of an internal and an external space in the classical
framework. The key problem is to identify the external spacetime
in which we are living. Many works appeal to the Anthropic
Principle [1]: there may exist five or more dimensions, however
only in the 4-dimensional nearly flat spacetime we, the observers,
would be able to exist.

In this letter we shall argue that, in the framework of quantum
cosmology, this problem can be solved in some toy models without
using the Anthropic Principle.

The quantum state of the universe is described by its wave
function $\Psi$. In the no-boundary universe [2], the wave
function is defined by the path integral over all compact
manifolds with the argument of the wave function as the only
boundary. The main contribution to the path integral comes from
the instanton solution. This is the so-called $WKB$ approximation.
Therefore,  the instanton can be thought as the seed of the
universe.

Let us study the following Freund-Rubin toy models [3]. The matter
content of the universe is an antisymmetric tensor field
$A^{\alpha_1 \dots \alpha_{s-1}}$ of rank $s-1$. Its field
strength is a completely antisymmetric tensor $F^{\alpha_1 \dots
\alpha_s}$. If $s=2$, then the matter field is Maxwell. The
Lorentzian action can be written as
\begin{equation}
I_{lorentz} = \frac{1}{16\pi} \int_M \left (R - 2\Lambda
-\frac{8\pi}{s}F^2 \right ) + \frac{1}{8\pi} \int_{\partial M} K,
\end{equation}
where $\Lambda$ is the cosmological constant, $R$ is the scalar curvature of
the spacetime $M$ and $K$ is the extrinsic curvature of its boundary $\partial M$.

The Einstein equation is
\begin{equation}
R^{\mu \nu} - \frac{1}{2} g^{\mu \nu} R + \Lambda g^{\mu \nu} =  8\pi  \theta^{\mu \nu},
\end{equation}
where the energy momentum tensor $\theta^{\mu \nu}$ is
\begin{equation}
\theta^{\mu \nu} = F_{\alpha_1 \dots \alpha_{s-1}}^{\;\;\;\;
\;\;\;\;\; \mu} F^{\alpha_1 \dots \alpha_{s-1} \nu} - \frac{1}{2s}
F_{\alpha_1 \dots \alpha_s}F^{\alpha_1 \dots \alpha_s}g^{\mu\nu}.
\end{equation}

The field equation is
\begin{equation}
g^{-1/2}\partial_\mu (g^{1/2} F^{\mu \alpha_2\dots \alpha_s})=0.
\end{equation}

We use indices $m, \dots$ for the manifold $M^s$ and $\bar{m},
\dots$ for $M^{d-s}$, respectively. We assume that $M^s$ and
$(M^{d-s})$ are  topologically spheres, and only components of the
field $F$ with all unbarred indices can be nonzero. From de Rham
cohomology, there exists unique harmonics in $S^s$ [4], i.e, the
solution to the field equation (4)
\begin{equation}
F^{\alpha_1 \dots \alpha_s} =  \kappa \epsilon^{\alpha_1 \dots
\alpha_s}(s!g_s)^{-1/2},
\end{equation}
where $g_s$ is the determinant of the metric of $M_s$, $\kappa$ is
a charge constant. We set $\kappa$ to be imaginary, for this
moment.

We first consider the case $\Lambda = 0$. From above one can
derive the scalar curvature for each factor space
\begin{equation}
R_s =  \frac{(n-s-1)8\pi \kappa^2 }{n-2}
\end{equation}
and
\begin{equation}
R_{n-s} = - \frac{(s-1)(n-s)8\pi \kappa^2}{s(n-2)}.
\end{equation}
It appears that the $F$ field behaves as a cosmological constant,
which is anisotropic with respect to the factor spaces.

The metrics of the factor spacetimes should be Einstein. The
created universe would select the manifolds with maximum symmetry.
This point can be justified in quantum cosmology . As we shall
show below, at the $WKB$ level, the relative creation probability
of the universe is exponential to the negative of the Euclidean
action of the seed instanton. The action is proportional to the
product of the volumes of the two factor manifolds. Maximization
of the volumes can be realized only by the manifolds with maximum
symmetries. Therefore, the instanton metric is a product of $S^s
\times S^{n-s}$. The metric signature of $S^s (S^{n-s})$ is
negative (positive) definite.  This is the instanton version of
the Freund-Rubin solution [3].

To obtain the Lorentzian spacetime, one can begin with the $S^s$
metric
\begin{equation}
ds^2_s = -dt^2 - \frac{\sin^2 (L_st)}{L_s^2}(d\chi^2 + \sin^2\chi
d\Omega^2_{s-2}),
\end{equation}
where $L_s$ is the radius of the $S^s$ and $d\Omega^2_{s-2}$
represents the unit $s-2-$sphere.

One can obtain the $s-$dimensional anti-de Sitter space by an
analytic continuation at a $s-1-$dimensional surface where the
metric is stationary. One can choose $\chi= \frac{\pi}{2}$ as the
surface, set $\omega = i(\chi - \frac{\pi}{2})$ and obtain the
metric with signature $(-, \dots,-,+)$
\begin{equation}
ds^2_s = -dt^2 - \frac{\sin^2 (L_st)}{L_s^2}(-d\omega^2 +
\cosh^2\omega d\Omega^2_{s-2}).
\end{equation}
Then one can analytically continue the metric through the null
surface at $t=0$ by redefining $\rho = \omega + \frac{i\pi}{2}$
and get the $s-$dimensional anti-de Sitter metric
\begin{equation}
ds^2_s = -dt^2 + \frac{\sin^2 (L_st)}{L_s^2}(d\rho^2 + \sinh^2\rho
d\Omega^2_{s-2}).
\end{equation}
The obtained Lorentzian spacetime is the product of the
$s-$dimensional anti-de Sitter space, which we consider as the
external spacetime, and a $S^{n-s}$, which is identified as the
internal space.  The apparent dimension of the spacetime is $s$
[3].

From the same $S^s$  one can also get a $s-$dimensional
hyperboliod by setting $\sigma = i(t - \frac{\pi}{2L_s})$
\begin{equation}
ds^2_s = d\sigma^2 + \frac{\cosh^2 (L_s\sigma )}{L_s^2}(d\rho^2 +
\sinh^2\rho d\Omega^2_{s-2}).
\end{equation}

One can also obtain the $n-s-$dimensional de Sitter space through
a simple analytic continuation from the factor space $S^{n-s}$ as
in the 4-dimensional case [2], and consider the positive definite
$s-$dimensional hyperboloid as the internal space. Then the
apparent dimension of the external spacetime becomes $n-s$ [3].

One can appeal to quantum cosmology to discriminate these two
possibilities. The relative creation probability of the universe
is
\begin{equation}
P =\Psi^* \cdot \Psi \approx \exp (-I) ,
\end{equation}
where $\Psi$ is the wave function of the configuration at the
quantum transition. The configuration is the metric and the matter
field at the equator. $I$ is the Euclidean action of the
instanton. It is worth emphasizing that the instanton is
constructed by joining its south hemisphere and its time reversal,
its north hemisphere.

In the Lorentzian regime, the probability of a quantum state is
independent of the representation. However, in the Euclidean regime
 this is not the case. In quantum cosmology, the universe is
created from nothing in imaginary time. In the Euclidean regime
the total relative probability of finding the universe  does not
stay constant. In fact, formula (12) can only be meaningful when
one uses a right representation for the wave function at the
equator. This problem was hidden in the earlier years of research
of quantum cosmology. At that stage, only regular instantons were
considered as seeds of universe creations.

Now, it is well known that regular instantons are too rare for the
creation scenario of a more realistic cosmological model. One has
to appeal to the constrained instantons [5]. The right
representation can be obtained through a canonical transform from
the wrong representation. The wave function subjects to a Fourier
transform in the Lorentzian regime. At the $WKB$ level, this
corresponds to a Legendre transform, the Legendre term at the
equator will change the probability value in Eq. (12). For a
regular instanton, one member of any pair of canonical conjugate
variables must vanish at the equator, so does the Legendre term.

The criterion for the right representation in formula (12) with a
constrained instanton is that across the equator the arguments of
the wave function must be continuous. This problem was encountered
in the problem  of quantum creation of magnetic and electric black
holes [6].  If one considers the quantum creation of a general
charged and rotating black hole, this point is even more critical.
It becomes so acute that unless the right configuration is used,
one cannot even find a constrained instanton seed [7].

Now, the action (1) is given under the condition that at the
boundary $\partial M$ the metric and the tensor field $A^{\alpha_1
\dots \alpha_{s-1}}$ are given. If we assume the external space is
the $s$-dimensional anti-de Sitter space, then the Euclidean
action is
\begin{equation}
I = \frac{1}{16\pi} \int_M \left (R - 2\Lambda -\frac{8\pi}{s}F^2
\right ) + \frac{1}{8\pi} \int_{\partial M} K,
\end{equation}
where all  quantities are Euclidean and the path of the
continuation from the Lorentzian action to the Euclidean action
has been such that the sign in front of $R$ term should be
positive. Since $R = R_{n-s} + R_s$, the negative value of $R_s$
is crucial for the perturbation calculation around the background
of the external spacetime. The right sign is necessary for the
primordial fluctuations to take the minimum excitation states
allowed by the Heisenberg Uncertainty Principle [8].

The action of the instanton can be evaluated as
\begin{equation}
I = \left (\frac{n- 2s}{2s(n-2)} - \frac{1}{2s} \right )\kappa^2
V_sV_{n-s},
\end{equation}
where the volumes $V_s$ and $V_{n-s}$ of $S^s$ and $S^{n-s}$ are
$2\pi^{(s+1)/2} L^s_s/\Gamma ((s+1)/2)$ and $2\pi^{(n-s+1)/2}
L^{n-s}_{n-s}/\Gamma ((n-s+1)/2)$, respectively, and $L_{n-s}$ is
the radius of the $S^{n-s}$.

The action is invariant under the gauge transformation
\begin{equation}
A_{\alpha_1
\dots \alpha_{s-1}} \longrightarrow A_{\alpha_1
\dots \alpha_{s-1}} + \partial_{[\alpha_1}\lambda_{\alpha_2
\dots \alpha_{s-1}]}.
\end{equation}
One can select a gauge such that there is only one nonzero
component  $A^{2 \dots s}$, where the index $1$ associated with
the time coordinate is excluded. There is no way to find a gauge
in which the field $A_{2 \dots s}$ is regular for the whole
manifold $S^s$ using single neighborhood. One can integrate (5) to
obtain its value at the equator with the regular condition at the
south pole. The field for the north hemisphere can be obtained
from the south solution through time reversal and a sign change.
This results in a discontinuity across the equator. When we
calculate the wave function of the universe, we implicitly fixed
the gauge and no freedom is left for the gauge transform. On the
other hand, the field strength $F^{\alpha_1 \dots \alpha_s}$ or
the canonical momentum $P^{2 \dots  s}$ is well defined and
continuous. Therefore, the field strength is the right
representation.

One can Fourier transform the wave function $\Psi(h_{ij}, A_{2
\dots s})$ to get the wave function $\Psi(h_{ij}, P^{2
\dots s})$
\begin{equation}
\Psi(h_{ij}, P^{2
\dots s}) = \frac{1}{2\pi} \int_{-\infty}^{\infty} e^{iA_{2
\dots s}P^{2
\dots s}}\Psi(h_{ij}, A_{2
\dots s}),
\end{equation}
where $h_{ij}$ is the metric of the equator. Here $A_{2 \dots s}$
is the only degree of freedom of the matter content under the
minisuperspace ansatz. $P^{2 \dots s}$ is defined as
\begin{equation}
P^{2 \dots s} = - \int_\Sigma (s-1)! F^{1 \dots s},
\end{equation}
where $\Sigma$ denotes the equator.

At the $WKB$ level, the Fourier transform is reduced into a
Legendre transform for the action. The Legendre transform
introduces an extra term $-2A_{2 \dots s}P^{2 \dots s}$ to the
Euclidean action $I$, where $A_{2 \dots s}$ is evaluated at the
south side of the equator. The two sides of the equator is taken
account by the factor $2$ here. The above calculation is carried
out for the equator $t =\frac{\pi}{2L_s}$. However the true
quantum transition should occur at $\chi = \frac{\pi}{2}$. Since
these two equators are congruent, the result should be the same.
This has also been checked.

It turns out the extra term is
\begin{equation}
I_{legendre} = \frac{1}{s} V_sV_{n-s} \kappa^2.
\end{equation}
Then the total action becomes
\begin{equation}
I_s = \left (\frac{n-2s}{2s(n-2)} + \frac{1}{2s} \right )\kappa^2
V_sV_{n-s}.
\end{equation}

Now if one uses the same instanton, and analytically continues
from the factor space $S^{n-s}$ at its equator to obtain an
$n-s-$dimensional de Sitter spacetime, and the internal space is
an $s-$dimensional hyperboloid. Then one still encounters the
representation problem of $A_{2 \dots s}$. In the context of our
argument, it has been implicitly assumed that for the argument of
the wave function the gauge is fixed. The singularity or
discontinuity is not avoidable. This is compatible with the fact
that the instanton is constrained. We know regular instantons are
either discrete or of constant action [5]. The action does depend
on the parameter $\kappa$ (see (20) below), therefore the
instanton does not qualify as a regular instanton. However, the
canonical momentum is zero here, and so is the Legendre term.

By the same argument as earlier for the continuation of the factor
space $S^s$, the Euclidean action should take an extra negative
sign, and the total action should be the negative of that in (14),
\begin{equation}
I_{n-s} =- \left (\frac{n-2s}{2s(n-2)} - \frac{1}{2s} \right
)\kappa^2 V_sV_{n-s}.
\end{equation}

From (12), we know the relative creation probability is the
exponential to the negative of the Euclidean action, therefore if
$2s-n <0$, then the creation probability of the universe with the
$s-$dimensional external space  exponentially dominates that with
the $n-s-$dimensional one, that is the apparent dimension is most
likely to be $s$. Otherwise the apparent dimension should be
$n-s$. If $2s = n$, then the two possibilities of creations are
equally likely.

One may also discuss the case with a real $\kappa$. For the case
$2s-n<0$, the universe is a product of an $s-$dimensional de
Sitter space and a $n-s-$dimensional hyperboloid. For the case $2s
-n>0$, the universe is a product of a $n-s-$dimensional anti-de
Sitter space and an $s-$dimensional sphere.

It is noted that the dimension of the external spacetime can never
be higher than that of the internal space.

In the $d=11$ supergravity, under a special ansatz, one can derive
the Freund-Rubin model with $n=11, s =4$.  It has been shown that
the apparent dimension must be 4 [9].

At this moment, it is instructive to recall the representation
problem in quantum creation of a Reissner-Nordstr$\rm\ddot{o}$m-de
Sitter black hole. In the ``regular" instanton case, the space is
the product $S^2 \times S^2$. The situation can be considered as a
special case of the Freund-Rubin toy models with $n=4, s=2$. If
the Maxwell field lives in the internal space, then we obtain a
magnetic black hole. For this case the charge, or $\kappa$, is
real. If the Maxwell field lives in the exterior space (or
2-dimensional de Sitter space in the Lorentzian regime), then the
black hole is electric. For this case, the charge, or $\kappa$, is
imaginary. As we mentioned, the electric or magnetic instanton is
not a regular instanton, it is a constrained instanton, therefore
one has to use a right representation as we explained above. In
the magnetic case the Legendre term is zero. After the Legendre
transform, the duality between the electric and magnetic black
holes is recovered, as far as the creation probability is
concerned [6][7].

\vspace*{0.1in}

\bf References:

\vspace*{0.1in}
\rm

1. S.W. Hawking, \it The Universe in a Nutshell  \rm (Bantam
Books, New York) chap 3 (2001).

2. J.B. Hartle and S.W. Hawking, \it Phys. Rev. \rm \bf D\rm
\underline{28} 2960 (1983).

3. G.O. Freund and M.A. Rubin,  \it Phys. Lett. \bf
B\rm\underline{97} 233 (1980).

4. T. Eguchi, P.B. Gilkey and A.J. Hanson \it Phys. Rep.
\rm\underline{66} 213 (1980).

5. Z.C. Wu, \it Gene. Rel. Grav.  \rm \underline{30} 1639 (1998),
hep-th/9803121.

6. S.W. Hawking and S.F. Ross, \it Phys. Rev. \rm \bf
D\rm\underline{52} 5865 (1995). R.B. Mann  and S.F. Ross , \it
Phys. Rev. \rm \bf D\rm\underline{52} 2254 (1995).

7. Z.C. Wu, \it Int. J. Mod. Phys. \rm \bf D\rm \underline{6} 199
(1997), gr-qc/9801020. Z.C. Wu, \it Phys. Lett. \bf B\rm
\underline{445} 274 (1999), gr-qc/9810012.

8. J.J. Halliwell and S.W. Hawking, \it Phys. Rev. \rm \bf
D\rm\underline{31}, 346 (1985).

9. Z.C. Wu, \it Phys. Rev. \rm\bf D\rm\underline{31}, 3079 (1985).
Z.C. Wu,  \it Gene. Rel. Grav.  \rm \underline{34} 1121 (2002),
hep-th/0105021.

\end{document}